\documentclass[aps,prl,reprint,balancelastpage,nofootinbib,preprintnumbers,superscriptaddress]{revtex4-1}

\usepackage{slashed}
\usepackage{bm}
\usepackage{latexsym,amssymb,amsmath,float,url,mathrsfs}
\usepackage{latexsym}
\usepackage{graphicx}
\usepackage{epstopdf}
\usepackage{amsmath}
\usepackage{comment}
\usepackage{subfigure}
\usepackage{natbib}
\usepackage{hyperref}
\usepackage{pifont}
\usepackage{blindtext}
\usepackage{adjustbox}
\usepackage{multirow}
\usepackage{tabularx}
\usepackage[table]{xcolor}
\usepackage{orcidlink}
\usepackage{tikz}
\usetikzlibrary{tikzmark}
\usepackage{fontawesome}

\usepackage{blkarray}
\usepackage{graphicx}
\usepackage{amsfonts}
\usepackage{soul}
\usepackage{amssymb}
\usepackage{amsmath}
\usepackage{cancel}
\usepackage{tcolorbox}

\usepackage{graphics,appendix,afterpage,makecell} 

\def\dd{{\rm d}}

\def\CC{C_{{\text{\tiny com}}}^{{\text{\tiny c}}}}
\def\CRP{C_{{\text{\tiny rp}}}}

\definecolor{oucrimsonred}{rgb}{0.6, 0.0, 0.0}
\definecolor{persianblue}{rgb}{0.11, 0.22, 0.73}
\definecolor{forestgreen}{rgb}{0.13,0.35,0.13}
\definecolor{lightgray}{rgb}{0.83, 0.83, 0.83}
 \hypersetup{colorlinks, citecolor=oucrimsonred, linkcolor=black, urlcolor=oucrimsonred}
\definecolor{cornellred}{rgb}{0.7, 0.11, 0.11}
\definecolor{navyblue}{rgb}{0.0, 0.0, 0.5}
\definecolor{amethyst}{rgb}{0.6, 0.4, 0.8}
\definecolor{yellow}{rgb}{1.0, 1.0, 0.0}
\definecolor{firebrick}{rgb}{0.7, 0.13, 0.13}
\definecolor{tangerineyellow}{rgb}{1.0, 0.8, 0.0}
\definecolor{deepfuchsia}{rgb}{0.76, 0.33, 0.76}
\definecolor{amber}{rgb}{1.0, 0.75, 0.0}
\definecolor{VioletRed4}{rgb}{0.55, 0.13, .32}
\definecolor{indiagreen}{rgb}{0.07, 0.53, 0.03}
\definecolor{VioletRed4}{rgb}{0.55, 0.13, .32}
\newcommand{\be}{\begin{equation}}
\newcommand{\ee}{\end{equation}}
\newcommand{\bea}{\begin{equation} \begin{aligned}}
\newcommand{\eea}{\end{aligned} \end{equation}}

\definecolor{oucrimsonred}{rgb}{0.6, 0.0, 0.0}
\newcommand\vertarrowbox[3][6ex]{%
  \begin{array}[t]{@{}c@{}} #2 \\
  \left\uparrow\vcenter{\hrule height #1}\right.\kern-\nulldelimiterspace\\
  \makebox[0pt]{\scriptsize#3}
  \end{array}%
}

\definecolor{verdechiaro}{rgb}{0.6,1,0.6}
\definecolor{giallochiaro}{rgb}{1,1,0.6}
\definecolor{bluscuro}{rgb}{0.15, 0.2, 0.9}
\definecolor{verdes}{rgb}{0.1, 0.5, 0.1}%
\definecolor{tangerineyellow}{rgb}{1.0, 0.8, 0.0}

\definecolor{americanrose}{rgb}{1.0, 0.01, 0.24}
\definecolor{cobalt}{rgb}{0.0, 0.28, 0.67}
\definecolor{brandeisblue}{rgb}{0.0, 0.44, 1.0}
\definecolor{mycolor}{rgb}{0.0, 0.0, 0.5}
\definecolor{oxfordblue}{rgb}{0.0, 0.13, 0.28}
\definecolor{azure}{rgb}{0.0, 0.5, 1.0}
\definecolor{turquoiseblue}{rgb}{0.0, 1.0, 0.94}
\newtcolorbox{mynewbox}[1]{colback=white!5!white,colframe=azure!75!black,fonttitle=\bfseries,title=#1}
\newtcolorbox{mybox}{colback=mycolor!5!white,colframe=azure!75!black}
\newtcolorbox{mynamedbox}[1]{colback=mycolor!5!white,colframe=azure!75!black,title=#1}
\definecolor{venetianred}{rgb}{0.78, 0.03, 0.08}
\newtcolorbox{mynamedbox1}[1]{colback=venetianred!5!white,colframe=venetianred!80!black,title=#1}
\newtcolorbox{mynamedbox2}[1]{colback=azure!5!white,colframe=azure!80!black,title=#1}

\definecolor{verdes}{rgb}{0.1, 0.5, 0.1}%
\definecolor{cornellred}{rgb}{0.7, 0.11, 0.11}

\definecolor{VioletRed4}{rgb}{0.55, 0.13, .32}

\hypersetup{
     colorlinks   = true,
     citecolor    = violet,
     urlcolor     = violet,
     linkcolor    = violet}

\definecolor{rossocorsa}{rgb}{0.83, 0.0, 0.0}

\usepackage[normalem]{ulem}

\newcommand{\papertitle}{The  Black Hole   Formation --  Null Geodesic Correspondence}

\begin{document}

\title[]{\papertitle}
\author{Andrea Ianniccari\orcidlink{0009-0008-9885-7737}}
\affiliation{Department of Theoretical Physics and Gravitational Wave Science Center,  \\
24 quai E. Ansermet, CH-1211 Geneva 4, Switzerland}

\author{Antonio J. Iovino\orcidlink{0000-0002-8531-5962}}
\affiliation{Department of Theoretical Physics and Gravitational Wave Science Center,  \\
24 quai E. Ansermet, CH-1211 Geneva 4, Switzerland}
\affiliation{Dipartimento di Fisica, ``Sapienza'' Universit\`a di Roma, Piazzale Aldo Moro 5, 00185, Roma, Italy}
\affiliation{INFN Sezione di Roma, Piazzale Aldo Moro 5, 00185, Roma, Italy}

\author{Alex Kehagias\orcidlink{}}
\affiliation{Physics Division, National Technical University of Athens, Athens, 15780, Greece}

\author{Davide Perrone\orcidlink{0000-0003-4430-4914}}
\affiliation{Department of Theoretical Physics and Gravitational Wave Science Center,  \\
24 quai E. Ansermet, CH-1211 Geneva 4, Switzerland}

\author{Antonio Riotto\orcidlink{0000-0001-6948-0856}}
\affiliation{Department of Theoretical Physics and Gravitational Wave Science Center,  \\
24 quai E. Ansermet, CH-1211 Geneva 4, Switzerland}


\begin{abstract}
\noindent
 We provide evidence for a correspondence between the formation of black holes and the stability of circular null geodesics around the collapsing perturbation. We first show that the critical threshold of the compaction function to form a black hole in radiation  is well approximated  by the critical threshold for the appearance of the first unstable circular orbit in a spherically symmetric background. We also show that  the critical exponent in the scaling law of the  primordial black hole mass close to the threshold  is set  by the inverse of the  Lyapunov coefficient of the unstable orbits when a self-similar stage is developed  close to criticality.

\end{abstract}

\maketitle

\noindent\textbf{Introduction --} 
Geodesic motions are crucial in determining the  fundamental features of spacetime.  Circular geodesics are particularly  interesting in this regard. For instance, the  binding energy of the last stable circular time-like geodesic in the Kerr geometry may be used to give an estimate  of  the spin of astrophysical black holes \cite{Zhang:1997dy,Narayan:2005ie,Shapiro}. 
Null unstable geodesics are  also intimately  linked  to the appearance of black holes to   external observers  and have been associated to the characteristic quasi-normal modes of black holes \cite{Press,Nollert_1999,Kokkotas:1999bd}
which  can be thought of as  null particles trapped at the unstable circular orbit and slowly leaking out
\cite{Goebel,Ferrari:1984zz,Mashhoon:1985cya,Berti:2005eb}. 
The real part of the quasi-normal
frequency is set  by the angular velocity at the
unstable null geodesic, while  the imaginary part has been shown to be  related to
the instability timescale of the orbit  \cite{Cornish:2003ig,Cardoso:2008bp}. Such a time scale is set by 
the Lyapunov exponent characterizing the rate of separation of infinitesimally close trajectories. 

Unstable circular orbits might also help to describe  phenomena occurring at the threshold of black hole formation in the high-energy scattering of black holes \cite{Pretorius:2007jn}.
Finally, there seems to be  a correspondence between the  scaling exponent setting   the number of orbits of two Schwarzschild black holes before merging into a Kerr black hole and  the Lyapunov coefficient of the circular orbit  geodesics of the final state Kerr black hole \cite{Pretorius:2007jn}, as if the properties of the null geodesics of the  final state are connected to  the dynamics leading to it.

In this letter we would like to build upon these results and propose some evidence of a correspondence between the formation of Black Holes (BHs), specifically in the radiation phase of the early universe and the properties of the null geodesics around the perturbation which eventually collapse into the BH final state. 

We will focus in the radiation phase as we will think of BHs 
formed in the early universe, the so-called Primordial Black Holes (PBHs). Indeed, 
they have become a focal point of interest in cosmology in recent years. In the standard scenario PBHs are formed by the gravitational collapse of sizeable perturbations generated during inflation (see Ref. \cite{LISACosmologyWorkingGroup:2023njw} for a recent review). However, our logical path following the physics of null geodesics can be applied to BHs formed in different environments and/or from different fields.

By characterizing the initial perturbation with the corresponding  compaction function, we will show that -- varying its  amplitude  -- the  critical value for  which the first circular orbit appears  with vanishing Lyapunov coefficient  well reproduces the critical value for which a BH is formed.  Furthermore, 
the formation of BHs at criticality is subsequent to a self-similar evolution which results in a final mass following  a scaling law with a universal critical exponent \cite{Choptuik:1993,Evans:1994}. 
We will be able to identify such critical exponent  with the inverse of the Lyapunov coefficient of the unstable circular orbits during the self-similar stage of the collapse. Before launching ourselves into the technical aspects, let us set the stage in the next section.

\vskip 0.5cm
\noindent
\textbf{ Geodesics stability and Lyapunov exponent --} In order to investigate the physics of null geodesics and its relation to the formation threshold of BHs,  we find it convenient to work with the  metric  in the  radial gauge and polar slicing (which we will call from now on radial polar gauge). 

These coordinates are the  generalization of the Schwarszschild coordinates to the non-static and non-vacuum spacetime and have been routinely used in the numerical studies of the gravitational collapse resulting in the formation of  BHs \cite{Choptuik:1993,Evans:1994}. 
The metric reads

\be
\label{metric}
\dd s^2=-\alpha^2(r,t)\dd t^2+a^2(r,t)\dd r^2+r^2\dd\Omega^2.
\ee
Let us  consider a physics situation in which the time dependence may be neglected and   stationarity can be assumed. 
Null geodesics are determined  by the trajectories which move along the equatorial plane such that

\be
-\alpha^2\dot t^2+a^2\dot r^2+r^2\dot\phi^2=0,
\ee
where the dots indicate differentation with respect to the affine parameter and $\phi$ is the azimuthal angle. Because of the spherical symmetry, one has $\dot\phi^2=L^2/r^4$, where $L$ is the angular momentum. Similarly, stationarity gives $\dot t^2=E^2/\alpha^4$, where $E$ is the conserved energy. The equation of motion can be written as

\be
\label{eq:potential}
\dot r^2=-V(r)=-\frac{1}{a^2}\left(-\frac{E^2}{\alpha^2}+\frac{L^2}{r^2}\right).
\ee
A circular orbit at a given radius $r_c$ exists if 

\be
V(r_c)=V'(r_c)=0,
\ee
where the prime indicates differentation with respect to radial coordinate. These conditions impose, respectively

\begin{eqnarray}
    \frac{E^2}{L^2}&=& \frac{\alpha^2_c}{r^2_c},\\
    \label{con}
    1&=&r_c\frac{\alpha'_c}{\alpha_c},
\end{eqnarray}
where the subscript ${}_c$ means that the quantity in question is evaluated at the radius $r_c$ of the  circular null geodesic. 

If we  slightly perturb the orbit taking $r=r_c+\delta r$ and Taylor expand the potential, we   get

\be
\delta\dot r^2\simeq - \frac{1}{2}V''(r_c)(\delta r)^2.
\ee
Writing $\delta\dot r=(\partial \delta r/\partial t)\dot t$,
we obtain

\be
\delta r(t)=\delta r(0)\,e^{\lambda t},
\ee
where

\be
\lambda=\sqrt{\frac{-V''(r_c)}{2\dot t^2(r_c)}}=\frac{1}{a_c}\sqrt{-\alpha_c\alpha''_c}
\ee
is the Lyapunov coefficient which determines the time scale of the  the instability of the circular orbits against small perturbations. 

One fundamental point to notice is the following. Let us  write the condition (\ref{con}) as $g(r_c)=0$,
where 
\be
g(r)\equiv1-r\alpha'(r)/\alpha(r).
\ee
If we  take the energy associated to the potential for generic time-like orbits we notice that
\begin{equation}
    E^2 = \frac{\alpha}{g(r)} =\frac{\alpha^2}{1-r\alpha'/\alpha}
\end{equation}
which implies $g(r)>0$ from the condition that this conserved quantity is indeed the energy and it is real. Furthermore the condition of light-like orbits corresponds to the innermost time-like orbit at radius $r=r_c$, implying $g(r_c)=0$.
Since $g(r)$ is positive for time-like orbits, by changing a parameter (which will be identified in Eq. (\ref{par})  as the amplitude of the compaction function $A$), one meets the critical radius at which the orbit becomes light-like. The first time this happens is when the condition $g(r_c)=0$ is reached at the minimum of $g(r)$, i.e.
\begin{equation}
    g_c = g'_c =-r_c\frac{\alpha''_c}{\alpha_c}=0\to \lambda=0.
\end{equation}
Let us imagine to change the parameter $A$. Initially no circular orbits are found (red lines of Fig. \ref{Lyapunov}).  
The first   critical value $r_c$ is obtained  when the  minimum of the function $g(r)$ vanishes, which signals  the point where the Lyapunov coefficient vanishes. Further increasing the parameter $A$, the curve $g(r)$ vanishes for two critical radii  (blue lines) for which  unstable orbits exist (on the right of the minimum). 
The position of the stable and unstable orbits can be understood as follows. The potential $V(r)$ (\ref{eq:potential}) in this case goes to $+\infty$ for $r\to 0$, having a minimum closer to $r\to 0$ and a maximum further away as can be seen in Fig. \ref{fig:f2}. The depth of the minimum is related to the parameter $A$ and it is coincident with the maximum at threshold.
As the BH forms the potential will change shape, developing the usual divergence to $-\infty$ for small radii instead. This change of asymptotic behaviour will get rid of the closer minimum, so we can think of the depth of the minimum as the one related to the mass of the future BH, which at threshold gives exactly a zero mass BH, while the outer maximum could be thought as the one related to the Innermost Stable Circular Orbit (ISCO) of the forming BH.

\begin{figure}[hbt]
\centering
  \includegraphics[width=8cm]{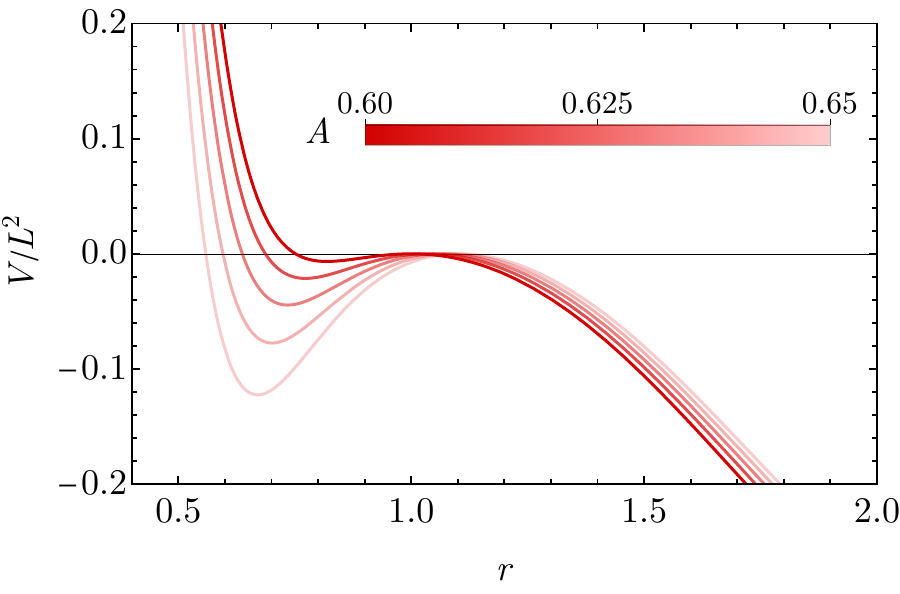}
  \caption{Plot of the potential obtained from Eq. (\ref{eq:potential}) for multiple values of the parameter $A$. At threshold we have no maximum or minimum, but an inflection point, while further increasing the values of $A$ give rise to a stable and an unstable circular orbit.}
  \label{fig:f2}
\end{figure}

\begin{figure}[hbt]
\centering
  \includegraphics[width=8cm]{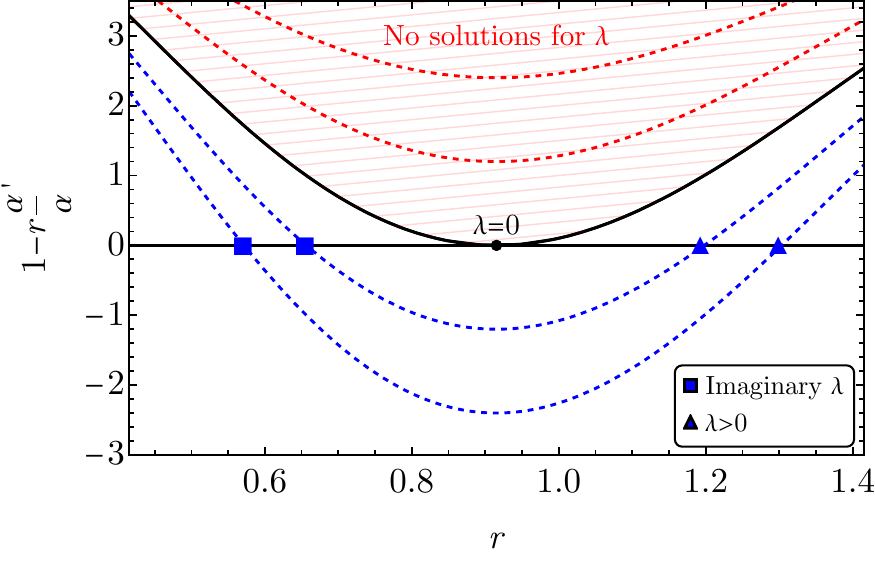}
  \caption{A schematic representation of the appearance of circular light-like orbits  
   by increasing amplitude of the compaction function from the red to the blue region.}
  \label{Lyapunov}
\end{figure}

\noindent\textbf{First evidence: the BH threshold from null geodesics --} 
Consider now a perturbation in the radiation energy density which re-enters the horizon after having been generated in a previous inflationary stage. Our fundamental  assumption is that in the first stage of the dynamics we may consider the fluid to be instantaneously at rest and we can neglect pressure gradients. This assumption  is supported  by  numerical simulations which show that maximum infall radial velocity remains  rather small  for a time considerably after horizon crossing, till  the perturbation has become highly non-linear \cite{Musco:2004ak}.\\
The compaction function in the radial polar gauge is\footnote{We use the subscript "rp"  for the radial polar gauge and "com" for the comoving gauge.}

\be
\CRP(r)=1- \frac{1}{a^2(r)}=\frac{8\pi G_N}{r}\int_0^r{\rm d}
x\, x^2\rho(x),
\ee
where $\rho(r)$ is the energy density perturbation. Combining the $(rr)$- and $(tt)$-Einstein equations, the lapse function satisfies the following equation  during the radiation phase   \cite{Evans:1994} 

\be\label{eq:ralph}
r\frac{\alpha'}{\alpha}=\frac{1}{6}\left(\frac{4\CRP+ r \CRP'}{1-\CRP}\right).
\ee
The condition for having a circular orbit therefore becomes

\be
\label{eq:crit_radius_2}
10\,\CRP(r_c)+ r_c \CRP'(r_c)=6.
\ee
This result is already encouraging as it provides values ${\cal O}(0.5)$ of the compaction function for which a circular orbit exists, that is  in the ballpark of the critical values for which we know BHs may form  \cite{LISACosmologyWorkingGroup:2023njw}.
The corresponding expression for the Lyapunov coefficient is

\be
\label{eq:lambda_special}
\lambda r_c=\alpha_c\sqrt{-\frac{1}{6}\left[11 r_c\CRP'(r_c)+r^2_c \CRP''(r_c)\right] },
\ee
which, for the argument of the previous section, vanishes  at the first value of  $r_c$ for which  Eq. (\ref{eq:crit_radius_2}) is satisfied.

The logic now is the following. The condition of vanishing Lyapunov coefficient selects a critical radius $r_c$, while the condition (\ref{eq:crit_radius_2}) selects the amplitude of the compaction function at that $r_c$. Larger values of the compaction function will have non-vanishing Lyapunov coefficients and therefore unstable circular orbits. 

Let us take as an example the compaction function of the form 
\be
\label{par}
 \CRP(r)=A_{{\text{\tiny rp}}}(r/r_0)^2 {\rm exp}\left[(1-(r/r_0)^{2k})/k\right],
\ee
which is the same as in Ref. \cite{Escriva:2019phb}, but in the radial polar gauge.
The condition $\lambda=0$ gives\footnote{We choose the $-$ branch of the square root because the other solution as $k\to 0$ gives a vanishing compaction function $\CRP(r_c)$.}

\be
\label{eq:crit_radius_3}
(r_c/r_0)^{2k}=\frac{1}{2}\left(7+k-\sqrt{25+14 k+k^2}\right),
\ee
which gives $r_c \sim 0.9$ $r_0$ for $k\simeq 0$ and $r_c \sim r_0$ for $k\gg 1$.
Imposing the condition (\ref{eq:crit_radius_2}) fixes the value of $A_{{\text{\tiny rp}}}$ and correspondingly of $\CRP(r_c)$, which turns out to be 
$\CRP(r_c)\simeq 0.6$ for $k\ll 1$ and 
$\CRP(r_c)\simeq0.5$ for $k\gg 1$.

Our goal is now to compare the maximum value of the compaction $\CRP(r)$ determined in this  way with the critical value of the compaction function to form a BH calculated numerically in the comoving gauge and on superhorizon scales and well  fitted by the formula \cite{Musco:2018rwt,Escriva:2019phb,Musco:2020jjb},

\be
\label{ff}
\CC(\widetilde r_m)=\frac{4}{15}e^{-1/q}\frac{q^{1-5/2q}}{\Gamma(5/2q)-\Gamma(5/2q,1/q)},
\ee
where $r_m$ is the location of the maximum of the compaction function and $q=- \widetilde r_m^2 C''_{{\text{\tiny com}}}(\widetilde r_m)/4 C_{{\text{\tiny com}}}(\widetilde r_m)$. 

To do so, we have to go from the radial polar gauge to the comoving gauge \cite{Harada:2015yda} defined with spatial coordinates $\widetilde r$
by knowing that the compaction function  is  coordinate invariant \cite{Misner:1964je}
\begin{figure}[hbt]
\centering
  \includegraphics[width=8cm]{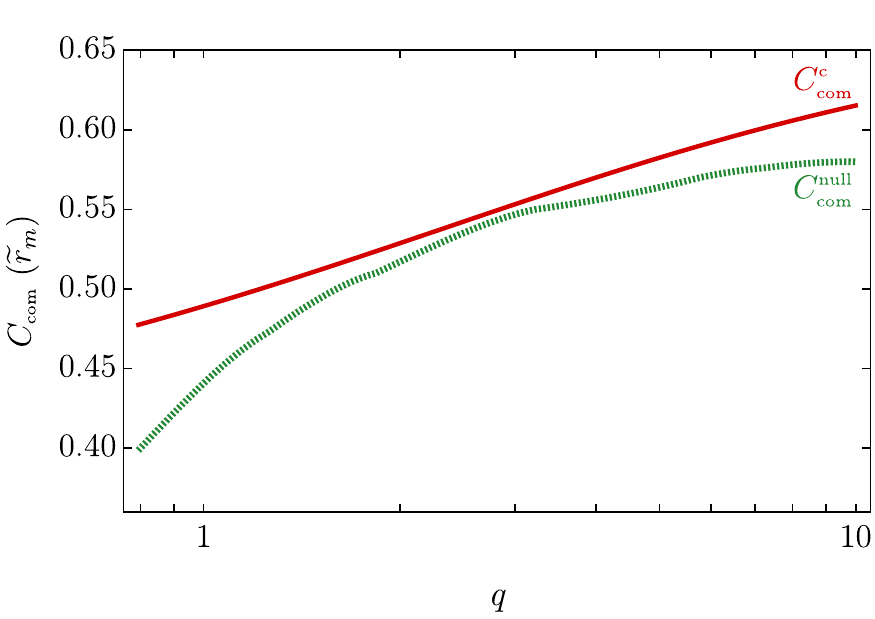}
  \caption{The comparison between the critical value of the compaction function from Refs. \cite{Musco:2018rwt,Escriva:2019phb,Musco:2020jjb} (red line) and the one obtained from the existence of the first circular orbit (dotted green line).}
  \label{critical}
\end{figure}
\be
C_{{\text{\tiny com}}}(\widetilde r)=C_{{\text{\tiny rp}}}(r(\widetilde{r})).
\ee
Explicitly
\begin{equation}\label{eq:Ccom}
C_{\mathrm{rp}}^{\text {null }}(r) \rightarrow C_{\mathrm{com}}^{\text {null }}(r)=1-\frac{1-C_{\mathrm{rp}}^{\text {null }}(r)}{\alpha(r)} .
\end{equation}
 By evaluating  it at  its maximum $\widetilde r_m(r_m)$ for which we have calculated the corresponding value of $q$, and by demanding the existence of the critical value $r_c$ such that $\lambda=0$ in the radial polar gauge, we obtain the critical compaction function in Fig. \ref{critical} (see the Supplementary Material
for more details). 
 
The  red solid line indicates the value (\ref{ff}), while the dotted green line indicates the compaction function in the comoving gauge corresponding to the appearance of the first unstable circular orbit. 
Admittedly, our correspondence fails for values of $q\ll 1$. Luckily, realistic models for BH formation do not have values in this regime. This is an impressive result given our  assumption  of neglecting the initial radial velocity. 
We have also checked that the result is stable against changing the parametrisation (\ref{par}). We also notice that the two critical values  depart more for  $q\gg 1$ as the threshold value from the expression (\ref{ff}) tends to $2/3\simeq 0.66$, while the one from the circular orbit reasoning increases up to $\sim 0.6$. This discrepancy  is not surprising as more peaked compaction functions are characterized by larger pressure gradients and our approximation is supposed to loose its validity in this regime.

\vskip 0.5cm

\noindent\textbf{Second evidence: the critical exponent  from null self-similar geodesics --} 
The gravitational collapse can be briefly described as follow. During its growth, when the comoving Hubble radius reaches the same size of a given overdensity, if the latter is larger than a critical  threshold,  a BH will form.
It is also the moment when the spacetime metric and the energy momentum tensor quickly approach a self-similar behaviour\,\cite{Choptuik:1993} which depends only on the variable
\be
z=\frac{r}{(-t)}, \quad t<0
\ee
and is independent from the time variable
\be
\tau=-\ln (-t)
\ee
At later times, self-similarity is broken, leading eventually to the formation of a $\mathrm{BH}$ if the evolution is super-critical, that is if the compaction function at its maximum is larger than a critical value (for a review, see Ref.\cite{Gundlach:2002sx}).
The resulting  BH mass follows a scaling relation of the type \cite{Choptuik:1993,Evans:1994,Musco:2008hv}
 
\be
\label{crit}
M_{\text{\tiny BH}}={\cal O}(1)M_{\text{\tiny H}}\left(C_{{\text{\tiny com}}}-C_{{\text{\tiny com}}}^{{\text{\tiny c}}}\right)^\gamma,\quad \gamma\simeq (0.35\div 0.37),
\ee
accounting for the mass of the BH  at formation written in units of the horizon mass $M_{\text{\tiny H}}$ at the time of horizon re-entry. The critical exponent $\gamma$ is universal 
reflecting a deep property of the gravitational dynamics. During the self-similar solution the dynamics depends only on the variable $z$ and not on the variable $\tau=-\ln (-t)$. The metric (\ref{metric}) is equivalent to

\begin{eqnarray}
\label{eq:metric1}
    \dd s^2 &=&  -f(z)\dd \tau^2 + a^2(z) \dd z^2 - 2 a^2(z) z \,\dd z \,\dd \tau + z^2  \dd\Omega^2,\nonumber\\
    f(z)&=& \alpha^2(z)  - z^2 a^2(z).
\end{eqnarray}
We can now repeat the same procedure as before to find the Lyapunov coefficient for the perturbed orbit around the critical ``radius" $z_c$ which satisfies the conditions
\begin{figure}[hbt]
\centering
  \includegraphics[width=8cm]{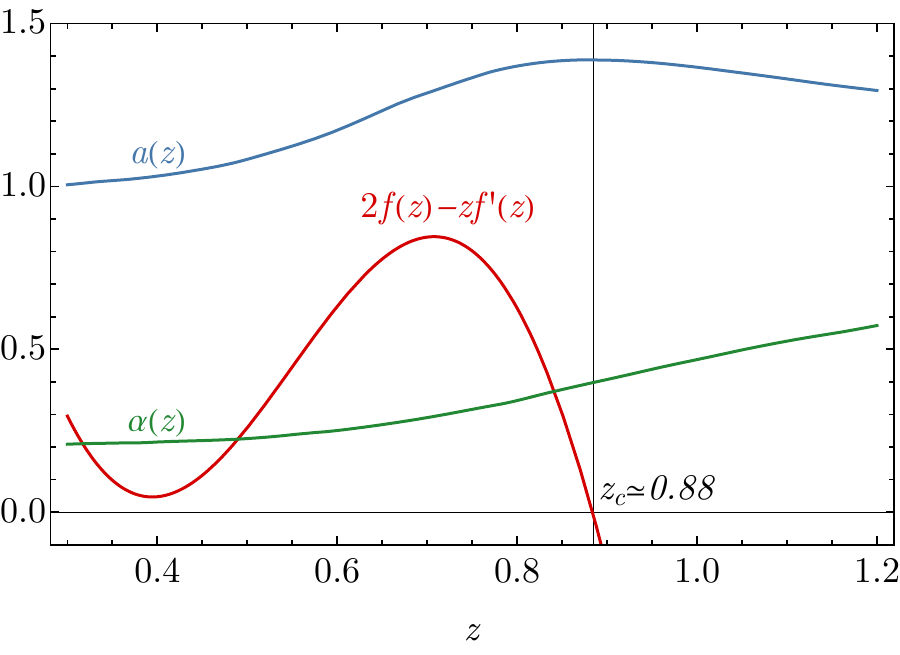}
  \caption{The values of the functions entering the Lyapunov coefficient in Eq. (\ref{eq:lyapunov_1}) reproducing the data from Ref. \cite{Evans:1994}. The chosen time is  $\tau=-0.5$, but the behaviour is self-similar, valid for all values of $\tau$ till self-similarity is broken at the formation of the BH. $z_c$ is determined by the condition $2f(z)-zf'(z)=0.$}
  \label{f}
\end{figure}
\vskip 0.5cm
\be
\frac{E^2}{L^2}=\frac{f_c}{z_c^2}\,\,\;{\rm and}\,\,\; 2f_c=z_c f'_c,
\ee
where now the prime indicates derivative with respect to $z$. 
The Lyapunov coefficient reads
\begin{eqnarray}
\label{eq:lyapunov_1}
    \lambda 
    &=& \frac{1}{\sqrt{2}}\sqrt{\frac {f_c}{z_c^2 \,\alpha_c^2 a_c^2} \left( 2f_c - z_c^2 f_c''\right)}
\end{eqnarray}
and determines the time scale of the unstable circular orbits

\be
\label{ly}
\delta z=\delta z_0 \,e^{\lambda \tau}.
\ee
We now note that  $\delta z_0$ will be proportional to $\left(C_{{\text{\tiny com}}}-C_{{\text{\tiny com}}}^{{\text{\tiny c}}}\right)$ for a family of geodesics that approach the unstable orbit when $C_{{\text{\tiny com}}}=C_{{\text{\tiny com}}}^{{\text{\tiny c}}}$. Perturbation theory breaks down when $\delta z\simeq 1$
which  sets the time when the geodesic will  depart from the circular orbit \cite{Pretorius:2007jn}. 
 We find from Eq. (\ref{ly})

\be
 \left(C_{{\text{\tiny com}}}-C_{{\text{\tiny com}}}^{{\text{\tiny c}}}\right)(-t)^{-\lambda}\sim 1.
\ee
On the other hand, the BH  mass $M_{\text{\tiny BH}}$ inside the apparent horizon is related to its radius by $M_{\text{\tiny BH}}= r_{\text{\tiny H}}/2G_N$. Replacing $-t_{\text{\tiny H}}=r_{\text{\tiny H}}/z_{\text{\tiny H}}$, we find

\be
M_{\text{\tiny BH}}\sim \left(C_{{\text{\tiny com}}}-C_{{\text{\tiny com}}}^{{\text{\tiny c}}}\right)^{1/\lambda}.
\ee
Now, the value of the Lyapunov coefficient can be extracted running the self-similar simulations following Ref. \cite{Evans:1994} (and whose results we will report elsewhere \cite{inprep}) and we have found  $\lambda\simeq 2.9$, giving $\gamma=1/\lambda\simeq 0.35$, which is very close to  the  value observed numerically in the literature \cite{Choptuik:1993,Evans:1994,Musco:2008hv}. 

An estimate to understand this result is the following. From Fig. \ref{f}, obtained by reproducing the results of Ref. \cite{Evans:1994}, one can appreciate that $z_c\simeq 1$ and  $f(z)\simeq -z^2 a^2(z)$. Under these approximations, the Lyapunov coefficient turns out to be $\lambda\sim (z_c^2/\alpha_c)\sqrt{a''_c a_c}$. One can estimate $a_c\simeq 1/\sqrt{1-\CRP}\simeq \sqrt{2}$ since $\CRP$ is always around  $0.5$. Similarly, $a_c''\simeq a_c/z_c^2$. Taking $\alpha_c\simeq 1/2$, the Lyapunov coefficient is
$\lambda\simeq 2\sqrt{2}\simeq 2.8$. This gives $\gamma=1/\lambda\simeq 0.36$,  which well approximates the numerical value.

\vskip 0.5cm
\vspace{5pt}\noindent\textbf{Further comments and conclusions  --} 
There is one more  piece of evidence of the correspondence we have proposed. Consider the moment when the BH has finally formed. Its mass follows the relation (\ref{crit}) with the critical exponent $\gamma\simeq 0.36$. For a BH, using the Schwarzschild metric, one easily finds that the circular orbit -- or photon ring -- exists at $r_c=3G_N M_{\text{\tiny BH}}$. Following the same logic to find the geodesics with $\alpha^2 = a^{-2} = 1- 2GM_{\text{\tiny BH}}/r$ in the metric, the corresponding Lyapunov coefficient is, in units of the horizon radius $r_{\text{\tiny H}}=2G_N M_{\text{\tiny BH}}$ and independently from the BH mass,
\begin{equation}
    \lambda =  \sqrt{\frac{f_c}{2r_c^2} (2f_c - r_c^2 f_c'')}, \quad f_c = 1-\frac{r_{\text{\tiny H}}}{r_c}
\end{equation}
which corresponds to
\be
\lambda r_{\text{\tiny H}}=\frac{2}{3\sqrt{3}}\simeq 0.38. 
\ee
The approximate equality  between this 
value and the value of the critical exponent  $\gamma$ (not its inverse) is striking. Furthermore, this (maybe only apparent) coincidence resembles the similarity -- which we have mentioned in the introduction --  between the  scaling exponent setting the number of orbits of two Schwarzschild BHs before merging into a Kerr BH and the Lyapunov coefficient of the circular orbit  geodesics of the final Kerr BH final state \cite{Pretorius:2007jn}. 
While we honestly do not have at the moment an understanding of such similarity, it will be interesting to investigate whether   it is  a further evidence of the correspondence between the formation of BHs and null geodesics. Other possible directions are  to check  if the correspondence works for other equations of state or other matter collapsing fields, {\it e.g.}   massless scalar fields.

\begin{acknowledgments}
\vspace{5pt}\noindent\emph{Acknowledgments --}
We thank V. De Luca for useful comments on the draft. A.I. and A.R.  acknowledge support from the  Swiss National Science Foundation (project number CRSII5\_213497).
A.J.I. acknowledges the financial support provided under the ``Progetti per Avvio alla Ricerca Tipo 1", protocol number AR1231886850F568.
D.~P. and A.R. are supported by the Boninchi Foundation for the project ``PBHs in the Era of GW Astronomy''.
\end{acknowledgments}

\newpage
\bibliography{Draft}

\begin{thebibliography}{25}%
\makeatletter
\providecommand \@ifxundefined [1]{%
 \@ifx{#1\undefined}
}%
\providecommand \@ifnum [1]{%
 \ifnum #1\expandafter \@firstoftwo
 \else \expandafter \@secondoftwo
 \fi
}%
\providecommand \@ifx [1]{%
 \ifx #1\expandafter \@firstoftwo
 \else \expandafter \@secondoftwo
 \fi
}%
\providecommand \natexlab [1]{#1}%
\providecommand \enquote  [1]{``#1''}%
\providecommand \bibnamefont  [1]{#1}%
\providecommand \bibfnamefont [1]{#1}%
\providecommand \citenamefont [1]{#1}%
\providecommand \href@noop [0]{\@secondoftwo}%
\providecommand \href [0]{\begingroup \@sanitize@url \@href}%
\providecommand \@href[1]{\@@startlink{#1}\@@href}%
\providecommand \@@href[1]{\endgroup#1\@@endlink}%
\providecommand \@sanitize@url [0]{\catcode `\\12\catcode `\$12\catcode `\&12\catcode `\#12\catcode `\^12\catcode `\_12\catcode `\%12\relax}%
\providecommand \@@startlink[1]{}%
\providecommand \@@endlink[0]{}%
\providecommand \url  [0]{\begingroup\@sanitize@url \@url }%
\providecommand \@url [1]{\endgroup\@href {#1}{\urlprefix }}%
\providecommand \urlprefix  [0]{URL }%
\providecommand \Eprint [0]{\href }%
\providecommand \doibase [0]{http://dx.doi.org/}%
\providecommand \selectlanguage [0]{\@gobble}%
\providecommand \bibinfo  [0]{\@secondoftwo}%
\providecommand \bibfield  [0]{\@secondoftwo}%
\providecommand \translation [1]{[#1]}%
\providecommand \BibitemOpen [0]{}%
\providecommand \bibitemStop [0]{}%
\providecommand \bibitemNoStop [0]{.\EOS\space}%
\providecommand \EOS [0]{\spacefactor3000\relax}%
\providecommand \BibitemShut  [1]{\csname bibitem#1\endcsname}%
\let\auto@bib@innerbib\@empty
\bibitem [{\citenamefont {Zhang}\ \emph {et~al.}(1997)\citenamefont {Zhang}, \citenamefont {Cui},\ and\ \citenamefont {Chen}}]{Zhang:1997dy}%
  \BibitemOpen
  \bibfield  {author} {\bibinfo {author} {\bibfnamefont {S.~N.}\ \bibnamefont {Zhang}}, \bibinfo {author} {\bibfnamefont {W.}~\bibnamefont {Cui}}, \ and\ \bibinfo {author} {\bibfnamefont {W.}~\bibnamefont {Chen}},\ }\href {\doibase 10.1086/310705} {\bibfield  {journal} {\bibinfo  {journal} {Astrophys. J. Lett.}\ }\textbf {\bibinfo {volume} {482}},\ \bibinfo {pages} {L155} (\bibinfo {year} {1997})},\ \Eprint {http://arxiv.org/abs/astro-ph/9704072} {arXiv:astro-ph/9704072} \BibitemShut {NoStop}%
\bibitem [{\citenamefont {Narayan}(2005)}]{Narayan:2005ie}%
  \BibitemOpen
  \bibfield  {author} {\bibinfo {author} {\bibfnamefont {R.}~\bibnamefont {Narayan}},\ }\href {\doibase 10.1088/1367-2630/7/1/199} {\bibfield  {journal} {\bibinfo  {journal} {New J. Phys.}\ }\textbf {\bibinfo {volume} {7}},\ \bibinfo {pages} {199} (\bibinfo {year} {2005})},\ \Eprint {http://arxiv.org/abs/gr-qc/0506078} {arXiv:gr-qc/0506078} \BibitemShut {NoStop}%
\bibitem [{\citenamefont {{Shapiro}}\ and\ \citenamefont {{Teukolsky}}(1983)}]{Shapiro}%
  \BibitemOpen
  \bibfield  {author} {\bibinfo {author} {\bibfnamefont {S.~L.}\ \bibnamefont {{Shapiro}}}\ and\ \bibinfo {author} {\bibfnamefont {S.~A.}\ \bibnamefont {{Teukolsky}}},\ }\href {\doibase 10.1002/9783527617661} {\emph {\bibinfo {title} {{Black holes, white dwarfs and neutron stars. The physics of compact objects}}}}\ (\bibinfo {year} {1983})\BibitemShut {NoStop}%
\bibitem [{\citenamefont {{Press}}(1971)}]{Press}%
  \BibitemOpen
  \bibfield  {author} {\bibinfo {author} {\bibfnamefont {W.~H.}\ \bibnamefont {{Press}}},\ }\href {\doibase 10.1086/180849} {\bibfield  {journal} {\bibinfo  {journal} {apjl}\ }\textbf {\bibinfo {volume} {170}},\ \bibinfo {pages} {L105} (\bibinfo {year} {1971})}\BibitemShut {NoStop}%
\bibitem [{\citenamefont {Nollert}(1999)}]{Nollert_1999}%
  \BibitemOpen
  \bibfield  {author} {\bibinfo {author} {\bibfnamefont {H.-P.}\ \bibnamefont {Nollert}},\ }\href {\doibase 10.1088/0264-9381/16/12/201} {\bibfield  {journal} {\bibinfo  {journal} {Classical and Quantum Gravity}\ }\textbf {\bibinfo {volume} {16}},\ \bibinfo {pages} {R159} (\bibinfo {year} {1999})}\BibitemShut {NoStop}%
\bibitem [{\citenamefont {Kokkotas}\ and\ \citenamefont {Schmidt}(1999)}]{Kokkotas:1999bd}%
  \BibitemOpen
  \bibfield  {author} {\bibinfo {author} {\bibfnamefont {K.~D.}\ \bibnamefont {Kokkotas}}\ and\ \bibinfo {author} {\bibfnamefont {B.~G.}\ \bibnamefont {Schmidt}},\ }\href {\doibase 10.12942/lrr-1999-2} {\bibfield  {journal} {\bibinfo  {journal} {Living Rev. Rel.}\ }\textbf {\bibinfo {volume} {2}},\ \bibinfo {pages} {2} (\bibinfo {year} {1999})},\ \Eprint {http://arxiv.org/abs/gr-qc/9909058} {arXiv:gr-qc/9909058} \BibitemShut {NoStop}%
\bibitem [{\citenamefont {{Goebel}}(1972)}]{Goebel}%
  \BibitemOpen
  \bibfield  {author} {\bibinfo {author} {\bibfnamefont {C.~J.}\ \bibnamefont {{Goebel}}},\ }\href {\doibase 10.1086/180898} {\bibfield  {journal} {\bibinfo  {journal} {apjl}\ }\textbf {\bibinfo {volume} {172}},\ \bibinfo {pages} {L95} (\bibinfo {year} {1972})}\BibitemShut {NoStop}%
\bibitem [{\citenamefont {Ferrari}\ and\ \citenamefont {Mashhoon}(1984)}]{Ferrari:1984zz}%
  \BibitemOpen
  \bibfield  {author} {\bibinfo {author} {\bibfnamefont {V.}~\bibnamefont {Ferrari}}\ and\ \bibinfo {author} {\bibfnamefont {B.}~\bibnamefont {Mashhoon}},\ }\href {\doibase 10.1103/PhysRevD.30.295} {\bibfield  {journal} {\bibinfo  {journal} {Phys. Rev. D}\ }\textbf {\bibinfo {volume} {30}},\ \bibinfo {pages} {295} (\bibinfo {year} {1984})}\BibitemShut {NoStop}%
\bibitem [{\citenamefont {Mashhoon}(1985)}]{Mashhoon:1985cya}%
  \BibitemOpen
  \bibfield  {author} {\bibinfo {author} {\bibfnamefont {B.}~\bibnamefont {Mashhoon}},\ }\href {\doibase 10.1103/PhysRevD.31.290} {\bibfield  {journal} {\bibinfo  {journal} {Phys. Rev. D}\ }\textbf {\bibinfo {volume} {31}},\ \bibinfo {pages} {290} (\bibinfo {year} {1985})}\BibitemShut {NoStop}%
\bibitem [{\citenamefont {Berti}\ and\ \citenamefont {Kokkotas}(2005)}]{Berti:2005eb}%
  \BibitemOpen
  \bibfield  {author} {\bibinfo {author} {\bibfnamefont {E.}~\bibnamefont {Berti}}\ and\ \bibinfo {author} {\bibfnamefont {K.~D.}\ \bibnamefont {Kokkotas}},\ }\href {\doibase 10.1103/PhysRevD.71.124008} {\bibfield  {journal} {\bibinfo  {journal} {Phys. Rev. D}\ }\textbf {\bibinfo {volume} {71}},\ \bibinfo {pages} {124008} (\bibinfo {year} {2005})},\ \Eprint {http://arxiv.org/abs/gr-qc/0502065} {arXiv:gr-qc/0502065} \BibitemShut {NoStop}%
\bibitem [{\citenamefont {Cornish}\ and\ \citenamefont {Levin}(2003)}]{Cornish:2003ig}%
  \BibitemOpen
  \bibfield  {author} {\bibinfo {author} {\bibfnamefont {N.~J.}\ \bibnamefont {Cornish}}\ and\ \bibinfo {author} {\bibfnamefont {J.~J.}\ \bibnamefont {Levin}},\ }\href {\doibase 10.1088/0264-9381/20/9/304} {\bibfield  {journal} {\bibinfo  {journal} {Class. Quant. Grav.}\ }\textbf {\bibinfo {volume} {20}},\ \bibinfo {pages} {1649} (\bibinfo {year} {2003})},\ \Eprint {http://arxiv.org/abs/gr-qc/0304056} {arXiv:gr-qc/0304056} \BibitemShut {NoStop}%
\bibitem [{\citenamefont {Cardoso}\ \emph {et~al.}(2009)\citenamefont {Cardoso}, \citenamefont {Miranda}, \citenamefont {Berti}, \citenamefont {Witek},\ and\ \citenamefont {Zanchin}}]{Cardoso:2008bp}%
  \BibitemOpen
  \bibfield  {author} {\bibinfo {author} {\bibfnamefont {V.}~\bibnamefont {Cardoso}}, \bibinfo {author} {\bibfnamefont {A.~S.}\ \bibnamefont {Miranda}}, \bibinfo {author} {\bibfnamefont {E.}~\bibnamefont {Berti}}, \bibinfo {author} {\bibfnamefont {H.}~\bibnamefont {Witek}}, \ and\ \bibinfo {author} {\bibfnamefont {V.~T.}\ \bibnamefont {Zanchin}},\ }\href {\doibase 10.1103/PhysRevD.79.064016} {\bibfield  {journal} {\bibinfo  {journal} {Phys. Rev. D}\ }\textbf {\bibinfo {volume} {79}},\ \bibinfo {pages} {064016} (\bibinfo {year} {2009})},\ \Eprint {http://arxiv.org/abs/0812.1806} {arXiv:0812.1806 [hep-th]} \BibitemShut {NoStop}%
\bibitem [{\citenamefont {Pretorius}\ and\ \citenamefont {Khurana}(2007)}]{Pretorius:2007jn}%
  \BibitemOpen
  \bibfield  {author} {\bibinfo {author} {\bibfnamefont {F.}~\bibnamefont {Pretorius}}\ and\ \bibinfo {author} {\bibfnamefont {D.}~\bibnamefont {Khurana}},\ }\href {\doibase 10.1088/0264-9381/24/12/S07} {\bibfield  {journal} {\bibinfo  {journal} {Class. Quant. Grav.}\ }\textbf {\bibinfo {volume} {24}},\ \bibinfo {pages} {S83} (\bibinfo {year} {2007})},\ \Eprint {http://arxiv.org/abs/gr-qc/0702084} {arXiv:gr-qc/0702084} \BibitemShut {NoStop}%
\bibitem [{\citenamefont {Bagui}\ \emph {et~al.}(2023)\citenamefont {Bagui} \emph {et~al.}}]{LISACosmologyWorkingGroup:2023njw}%
  \BibitemOpen
  \bibfield  {author} {\bibinfo {author} {\bibfnamefont {E.}~\bibnamefont {Bagui}} \emph {et~al.} (\bibinfo {collaboration} {LISA Cosmology Working Group}),\ }\href@noop {} {\  (\bibinfo {year} {2023})},\ \Eprint {http://arxiv.org/abs/2310.19857} {arXiv:2310.19857 [astro-ph.CO]} \BibitemShut {NoStop}%
\bibitem [{\citenamefont {Choptuik}(1993)}]{Choptuik:1993}%
  \BibitemOpen
  \bibfield  {author} {\bibinfo {author} {\bibfnamefont {M.~W.}\ \bibnamefont {Choptuik}},\ }\href {\doibase 10.1103/PhysRevLett.70.9} {\bibfield  {journal} {\bibinfo  {journal} {Phys. Rev. Lett.}\ }\textbf {\bibinfo {volume} {70}},\ \bibinfo {pages} {9} (\bibinfo {year} {1993})}\BibitemShut {NoStop}%
\bibitem [{\citenamefont {Evans}\ and\ \citenamefont {Coleman}(1994)}]{Evans:1994}%
  \BibitemOpen
  \bibfield  {author} {\bibinfo {author} {\bibfnamefont {C.~R.}\ \bibnamefont {Evans}}\ and\ \bibinfo {author} {\bibfnamefont {J.~S.}\ \bibnamefont {Coleman}},\ }\href {\doibase 10.1103/PhysRevLett.72.1782} {\bibfield  {journal} {\bibinfo  {journal} {Phys. Rev. Lett.}\ }\textbf {\bibinfo {volume} {72}},\ \bibinfo {pages} {1782} (\bibinfo {year} {1994})}\BibitemShut {NoStop}%
\bibitem [{\citenamefont {Musco}\ \emph {et~al.}(2005)\citenamefont {Musco}, \citenamefont {Miller},\ and\ \citenamefont {Rezzolla}}]{Musco:2004ak}%
  \BibitemOpen
  \bibfield  {author} {\bibinfo {author} {\bibfnamefont {I.}~\bibnamefont {Musco}}, \bibinfo {author} {\bibfnamefont {J.~C.}\ \bibnamefont {Miller}}, \ and\ \bibinfo {author} {\bibfnamefont {L.}~\bibnamefont {Rezzolla}},\ }\href {\doibase 10.1088/0264-9381/22/7/013} {\bibfield  {journal} {\bibinfo  {journal} {Class. Quant. Grav.}\ }\textbf {\bibinfo {volume} {22}},\ \bibinfo {pages} {1405} (\bibinfo {year} {2005})},\ \Eprint {http://arxiv.org/abs/gr-qc/0412063} {arXiv:gr-qc/0412063} \BibitemShut {NoStop}%
\bibitem [{\citenamefont {Escriv\`a}\ \emph {et~al.}(2020)\citenamefont {Escriv\`a}, \citenamefont {Germani},\ and\ \citenamefont {Sheth}}]{Escriva:2019phb}%
  \BibitemOpen
  \bibfield  {author} {\bibinfo {author} {\bibfnamefont {A.}~\bibnamefont {Escriv\`a}}, \bibinfo {author} {\bibfnamefont {C.}~\bibnamefont {Germani}}, \ and\ \bibinfo {author} {\bibfnamefont {R.~K.}\ \bibnamefont {Sheth}},\ }\href {\doibase 10.1103/PhysRevD.101.044022} {\bibfield  {journal} {\bibinfo  {journal} {Phys. Rev. D}\ }\textbf {\bibinfo {volume} {101}},\ \bibinfo {pages} {044022} (\bibinfo {year} {2020})},\ \Eprint {http://arxiv.org/abs/1907.13311} {arXiv:1907.13311 [gr-qc]} \BibitemShut {NoStop}%
\bibitem [{\citenamefont {Musco}(2019)}]{Musco:2018rwt}%
  \BibitemOpen
  \bibfield  {author} {\bibinfo {author} {\bibfnamefont {I.}~\bibnamefont {Musco}},\ }\href {\doibase 10.1103/PhysRevD.100.123524} {\bibfield  {journal} {\bibinfo  {journal} {Phys. Rev. D}\ }\textbf {\bibinfo {volume} {100}},\ \bibinfo {pages} {123524} (\bibinfo {year} {2019})},\ \Eprint {http://arxiv.org/abs/1809.02127} {arXiv:1809.02127 [gr-qc]} \BibitemShut {NoStop}%
\bibitem [{\citenamefont {Musco}\ \emph {et~al.}(2021)\citenamefont {Musco}, \citenamefont {De~Luca}, \citenamefont {Franciolini},\ and\ \citenamefont {Riotto}}]{Musco:2020jjb}%
  \BibitemOpen
  \bibfield  {author} {\bibinfo {author} {\bibfnamefont {I.}~\bibnamefont {Musco}}, \bibinfo {author} {\bibfnamefont {V.}~\bibnamefont {De~Luca}}, \bibinfo {author} {\bibfnamefont {G.}~\bibnamefont {Franciolini}}, \ and\ \bibinfo {author} {\bibfnamefont {A.}~\bibnamefont {Riotto}},\ }\href {\doibase 10.1103/PhysRevD.103.063538} {\bibfield  {journal} {\bibinfo  {journal} {Phys. Rev. D}\ }\textbf {\bibinfo {volume} {103}},\ \bibinfo {pages} {063538} (\bibinfo {year} {2021})},\ \Eprint {http://arxiv.org/abs/2011.03014} {arXiv:2011.03014 [astro-ph.CO]} \BibitemShut {NoStop}%
\bibitem [{\citenamefont {Harada}\ \emph {et~al.}(2015)\citenamefont {Harada}, \citenamefont {Yoo}, \citenamefont {Nakama},\ and\ \citenamefont {Koga}}]{Harada:2015yda}%
  \BibitemOpen
  \bibfield  {author} {\bibinfo {author} {\bibfnamefont {T.}~\bibnamefont {Harada}}, \bibinfo {author} {\bibfnamefont {C.-M.}\ \bibnamefont {Yoo}}, \bibinfo {author} {\bibfnamefont {T.}~\bibnamefont {Nakama}}, \ and\ \bibinfo {author} {\bibfnamefont {Y.}~\bibnamefont {Koga}},\ }\href {\doibase 10.1103/PhysRevD.91.084057} {\bibfield  {journal} {\bibinfo  {journal} {Phys. Rev. D}\ }\textbf {\bibinfo {volume} {91}},\ \bibinfo {pages} {084057} (\bibinfo {year} {2015})},\ \Eprint {http://arxiv.org/abs/1503.03934} {arXiv:1503.03934 [gr-qc]} \BibitemShut {NoStop}%
\bibitem [{\citenamefont {Misner}\ and\ \citenamefont {Sharp}(1964)}]{Misner:1964je}%
  \BibitemOpen
  \bibfield  {author} {\bibinfo {author} {\bibfnamefont {C.~W.}\ \bibnamefont {Misner}}\ and\ \bibinfo {author} {\bibfnamefont {D.~H.}\ \bibnamefont {Sharp}},\ }\href {\doibase 10.1103/PhysRev.136.B571} {\bibfield  {journal} {\bibinfo  {journal} {Phys. Rev.}\ }\textbf {\bibinfo {volume} {136}},\ \bibinfo {pages} {B571} (\bibinfo {year} {1964})}\BibitemShut {NoStop}%
\bibitem [{\citenamefont {Gundlach}(2003)}]{Gundlach:2002sx}%
  \BibitemOpen
  \bibfield  {author} {\bibinfo {author} {\bibfnamefont {C.}~\bibnamefont {Gundlach}},\ }\href {\doibase 10.1016/S0370-1573(02)00560-4} {\bibfield  {journal} {\bibinfo  {journal} {Phys. Rept.}\ }\textbf {\bibinfo {volume} {376}},\ \bibinfo {pages} {339} (\bibinfo {year} {2003})},\ \Eprint {http://arxiv.org/abs/gr-qc/0210101} {arXiv:gr-qc/0210101} \BibitemShut {NoStop}%
\bibitem [{\citenamefont {Musco}\ \emph {et~al.}(2009)\citenamefont {Musco}, \citenamefont {Miller},\ and\ \citenamefont {Polnarev}}]{Musco:2008hv}%
  \BibitemOpen
  \bibfield  {author} {\bibinfo {author} {\bibfnamefont {I.}~\bibnamefont {Musco}}, \bibinfo {author} {\bibfnamefont {J.~C.}\ \bibnamefont {Miller}}, \ and\ \bibinfo {author} {\bibfnamefont {A.~G.}\ \bibnamefont {Polnarev}},\ }\href {\doibase 10.1088/0264-9381/26/23/235001} {\bibfield  {journal} {\bibinfo  {journal} {Class. Quant. Grav.}\ }\textbf {\bibinfo {volume} {26}},\ \bibinfo {pages} {235001} (\bibinfo {year} {2009})},\ \Eprint {http://arxiv.org/abs/0811.1452} {arXiv:0811.1452 [gr-qc]} \BibitemShut {NoStop}%
\bibitem [{\citenamefont {Ianniccari}\ \emph {et~al.}(tion)\citenamefont {Ianniccari} \emph {et~al.}}]{inprep}%
  \BibitemOpen
  \bibfield  {author} {\bibinfo {author} {\bibfnamefont {A.}~\bibnamefont {Ianniccari}} \emph {et~al.},\ }\href@noop {} {\  (\bibinfo {year} {in preparation})}\BibitemShut {NoStop}%
\end{thebibliography}%
\clearpage
\newpage
\maketitle
\onecolumngrid
\begin{center}
\textbf{\large \papertitle} 
\\ 
\vspace{0.05in}
{Andrea Ianniccari, Antonio J. Iovino, Alex Kehagias, Davide Perrone and Antonio Riotto}
\\ 
\vspace{0.05in}
{ \it Supplementary Material}
\end{center}
\onecolumngrid
\setcounter{equation}{0}
\setcounter{figure}{0}
\setcounter{section}{0}
\setcounter{table}{0}
\setcounter{page}{1}
\makeatletter
\renewcommand{\theequation}{S\arabic{equation}}
\renewcommand{\thefigure}{S\arabic{figure}}
\renewcommand{\thetable}{S\arabic{table}}

\vspace{-0.5cm}
\section{Black hole threshold from null geodesic}\label{supp:1}
In this appendix we report step by step the procedure to get the values of the compaction function in the comoving gauge reported in eq.(\ref{eq:Ccom}) and showed in Fig.\,\ref{critical}.
\begin{itemize}
    \item \textit{I)} We fix a value of $k$,
    \item \textit{II)} we find the critical radius $r_c$ with Eq. (\ref{eq:crit_radius_3}) and the critical value of the amplitude $A_{\rm rp}$ through Eq. (\ref{eq:crit_radius_2}),
    \item \textit{III)} we evaluate $\alpha(r)$ through numerical integration fixing the condition $\alpha(r\to \infty)=1$. We compute analytically the r.h.s of Eq. (\ref{eq:ralph}) using the family profile in Eq. (\ref{par}) and we get
\be
\alpha(r)=\textrm{Exp}\left[-\int_r^{\infty} \frac{\dd x}{x} \frac{A_{\rm rp} e^{\frac{1}{k}} r^2\left(-3+r^{2 k}\right)}{-3 e^{\frac{r^{2 k}}{k}}+3 A_{\rm rp} e^{\frac{1}{k}} r^2} \right],
\ee

\item \textit{IV)} we use Eq. (\ref{eq:Ccom}) to obtain the compaction function in the comoving gauge $C_{\rm com}(r)$ and we evaluate its maximum point $r_m$ and its maximum $C_{\rm com}(r_m)$,

\item \textit{V)} we compute $q$ 
\be
q= - r_m^2 \frac{C_{\rm com}''(r_m)}{4 C_{\rm com}(r_m) }
\ee
using the new compaction by performing the second derivative evaluated at $r_m$, plotting it against $C_{\rm com}(r_m)$.

\end{itemize}

\end{document}